Brain and Cognitive Reserve: Translation via Network Control Theory

John Dominic Medaglia[1]

Fabio Pasqualetti[2]

Roy H. Hamilton[3]

Sharon L. Thompson-Schill[1]

Danielle S. Bassett[4,5]

1: Department of Psychology, University of Pennsylvania, Philadelphia, Pennsylvania 19104
2: Department of Mechanical Engineering, University of California-Riverside, Riverside, California 92521
3: Department of Neurology, Perelman School of Medicine, University of Pennsylvania, Philadelphia, Pennsylvania 19104
4: Department of Bioengineering, University of Pennsylvania, Pennsylvania 19104
5: Department of Electrical and Systems Engineering, University of Pennsylvania, Philadelphia, Pennsylvania 19104

Contact: Danielle S. Bassett
School of Engineering and Applied Sciences
Department of Bioengineering
240 Skirkanich Hall
210 South 33[rd] Street
Philadelphia, PA 19104-6321
dsb@seas.upenn.edu

**Abstract**

Traditional approaches to understanding the brain's resilience to neuropathology have identified neurophysiological variables, often described as brain or cognitive "reserve," associated with better outcomes. However, mechanisms of function and resilience in large-scale brain networks remain poorly understood. Dynamic network theory may provide a basis for substantive advances in understanding functional resilience in the human brain. In this perspective, we describe recent theoretical approaches from network control theory as a framework for investigating network level mechanisms underlying cognitive function and the dynamics of neuroplasticity in the human brain. We describe the theoretical opportunities offered by the application of network control theory at the level of the human connectome to understand cognitive resilience and inform translational intervention.

**Keywords**
Network science; neuropsychology; graph theory, control theory; neurology

**Introduction**

The brain is an intricately connected dynamic system that supports substantial information processing capacity underlying human thought (Marois & Ivanoff, 2005). How complex cognitive processes are executed in the brain remains a deeply challenging and unsolved question. Several recent lines of investigation suggest that healthy cognitive function relies on spatiotemporally interdependent (or *networked*) neurophysiological mechanisms: information transmission along white matter tracts, and neural computations within distributed networks of brain areas (cf. Kopell et al., 2014; Medaglia et al., 2015). In kind, abnormal cognitive function may depend on disruptions in networked mechanisms, altering the dynamic propagation of information and the healthy evolution of brain states (Da Silva et al., 2003; Pezard et al., 1996; Stam, 2014; van den Heuvel & Sporns, 2013). In the context of these emerging hypotheses, a major challenge remains in the development of generalized theories that account for cognitive function and dysfunction directly from neurophysiological mechanisms that operate at a network level..

Since the pioneering work of Hodgkin and Huxley in the 1940s and 50s, many approaches have been developed to address problems in neural dynamics at cellular and ensemble levels. Yet their implications for cognitive dysfunction in human disease remain largely unknown. Emerging techniques from the mathematical, physical, and engineering sciences may be able to address these challenges when applied to large-scale neuroimaging of the human brain. In particular, dynamic network theory offers an especially useful framework to examine networked mechanisms of brain function and dysfunction as it evolves during cognitive processes.

*Dynamic network theory* concerns how the time-evolving interactions between many interconnected elements result in complex system behavior. In applications to other real-world systems, techniques from this field have provided fundamental explanations for the emergence of complicated system dynamics from the interactions between system parts (Choi et al., 2001; Yamashita et al., 2008; Canard et al., 2012). Moreover, alterations in system function following perturbation or damage have been explained by the spread or diffusion of signals through the system's network (Albert et al., 2000; Boguna et al., 2003; Buldyrev et al., 2010). While these approaches have been developed in other contexts, the problems that they address are strikingly similar to the problem of explaining healthy and diseased cognitive processes using networked neurophysiological mechanisms. Should this similarity be more than a metaphor, the translation of these approaches to the cognitive and clinical neurosciences may prove crucial to addressing longstanding challenges in the brain and cognitive sciences (see Figure 1).

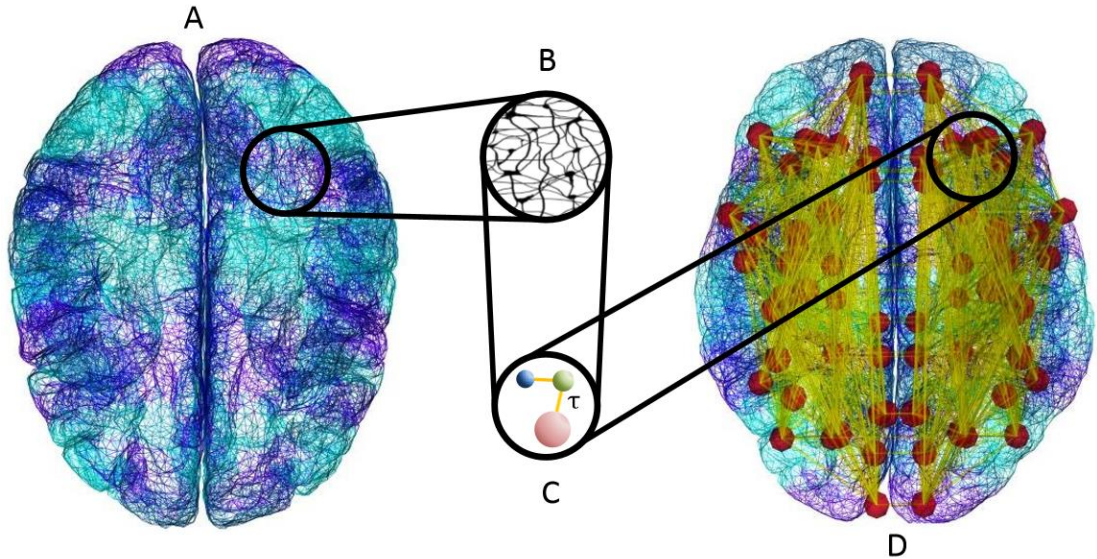

*Figure 1A-D:* Schematic representation of dynamic network theory in the human brain. (a) The brain can be separated into differentiable regions based upon cellular architecture (b) or systems containing functionally similar neurons. (c) Neural activities can be represented in a network representation consisting of nodes (spheres) and edges (connectors). Dynamics (τ) are represented along edges and are the activities important to healthy functioning. Nodes and edges that support dynamics can be represented at a coarse "macro" scale of brain organization (d), which is our current focus. Reproduced with permission and modified from (Leon et al., 2013).

Our goal is to understand large-scale functional properties of the human brain, how these properties support cognition, and under what conditions they fail. In clinical presentations, dynamic network theory posits that dysfunction is a result of aberrations in network dynamics. These aberrations can result from the disruption of network structures that support dynamics, the direct disruption of dynamics, or a mixture of the two. Indeed, conceptually, network pathways to disease may occur through structural failures in brain networks (Stam, 2014). These structural alterations may be complemented by alterations in neurophysiological dynamics that support brain function and cognition at multiple spatiotemporal resolutions (Kopell, 2014).

Dynamic network techniques offer two powerful advantages in understanding healthy cognition and its alteration in disease or injury. First, dynamic network approaches provide a basis for a formal union between mathematical approaches to complex systems and neurophysiological processes that support cognition. Mathematical axioms and analytic techniques from the emerging field of network science can enter the vocabulary and repertoire of the neurosciences. This affords the ability to conceptualize neuroscientific questions in a robust theoretical framework that has been progressively developing since the 1760s (Euler, 1766). As a result, the cognitive neuroscientist, neuropsychologist, and neurologist can enjoy and benefit from the quantitatively rigorous network representations of neuroimaging data, and directly probe their potential utility in uncovering fundamental insights into cognitive function in health and disease using empirical approaches.

Second, dynamic network approaches can be used to directly inform the manipulation of cognitive outcomes. As system dynamics and their generating network mechanisms are clarified, candidate targets for modification and repair can be proposed. This is particularly crucial to neurological and psychiatric diseases, where impairments in cognitive function are a primary concern in diagnosis and treatment. By drawing on developing methodologies in dynamic network theory, similarities between observed dysfunction in pathological syndromes and features in perturbed dynamic systems can be described. Initial interventions for the brain can be proposed based on the observed dynamic aberrations.

For the purposes of the current paper, we focus on one type of dynamic network analysis and describe its potential to inform theoretical and practical approaches to problems in cognitive dysfunction in neurological syndromes. *Network control theory* is an innovative and leading subfield of dynamic network theory that offers a class of powerful engineering-based conceptual and analytic approaches to examining functional signaling and resilience in networked systems. As a developing subfield, network control theory contains concepts that have been successfully applied to understand, manipulate, and repair complex systems in robotic, technological, and mechanical contexts. We suggest that these conceptual and practical approaches carry distinct advantages in developing brain connectomics into a translationally relevant field of study.

We briefly summarize key principles of network control theory and delineate their implications as an attractive approach to augment those typically taken in clinical neuroscience research, particularly in explaining brain and cognitive "reserve". Namely, we will emphasize the distinct advantage of a control-theoretic perspective on problems in brain structure, function, and cognition in neurological samples. To maintain clarity throughout this review, we consider brain structure and function to be measurable qualities of the brain's morphology and dynamics, respectively. Cognition is represented in the brain's structure and function, and its outputs are measurable in behavioral paradigms in experimental and clinical settings. After providing a basic introduction to reserve and network control theory in this context, we describe the application of network control theory to brain network structure and dynamics in the macro-scale human connectome (cf. Sporns, Tononi & Kotter, 2005).

We close with a speculative discussion of immediate extensions of network control theory to theoretical and analytic issues in understanding cognitive resilience in neurological diseases and implications for informing treatments. We provide initial hypotheses within this framework. We consider the potential for dynamic network approaches to introduce a conceptual framework for understanding variance in clinical trajectories and to delineate novel features of disease syndromes and targets for translational interventions. Crucially, we suggest that while this area is in its earliest stages, it carries the correct ingredients to promote productive scientific inquiry as the tools from several fields are sufficiently maturing.

**A Definition of Reserve**

In the clinical cognitive neurosciences, the constructs of "brain reserve" and "cognitive reserve" have been invoked to explain the imperfect correlation between brain pathology and clinical sequelae in numerous brain disorders (Stern, 2002). Generally, reserve (of either type) represents individual variability in the functional use or structural integrity of the nervous system that alters a person's cognitive and behavioral abilities following the onset of brain pathology. In essence, the concept of reserve suggests that some initial conditions of brain physiology and function – often measured using neuroimaging techniques– heavily constrain the observed clinical sequelae. Applied most extensively in the context of dementia but with recent applications in contexts such as brain injury, "reserve" is frequently used as a placeholder for a more precise understanding of the mechanisms of resilience in the face of neuropathology. A particular emphasis is placed on the variability in cognitive outcomes in spite of superficially similar effects of disease or damage to the brain, such as damage to gray or white matter structures in brain injury.

One can describe cognitive outcomes in the language of engineering by noting that the initial conditions of brain pathology constrain the *trajectory* of brain function through disease. The study of the relationship between initial conditions and nervous system trajectories is the principle goal of a classical branch of engineering known as dynamic systems theory. Yet, a direct link between the clinical cognitive understanding of reserve and systems science that might explain mechanisms of reserve remains underexplored. Here we posit that improved theoretical models for these links will advance our understanding of resilience in brain diseases and provide the basis for innovative translational approaches.

We focus on two commonly studied types of reserve: brain reserve and cognitive reserve. *Brain reserve* is a "passive" form of capacity that is thought to depend on the structural properties of the brain. Patients with less brain reserve are thought to have a lower threshold for the expression of functional impairments following the onset of brain pathology. A specific hypothesis of this model is that as brain volume or synaptic density decreases, individuals with more premorbid brain reserve will express symptoms more slowly and less severely than individuals with less premorbid brain reserve. This model has found significant support in cohorts with Alzheimer's disease (Graves et al., 1996; Murray et al., 2011; Perneczky et al., 2010; Satz, 1993; Schofield et al., 1997; Stern, 2002).

*Cognitive reserve* describes an "active" mechanism for coping with brain pathology (Stern, 2002). In contrast to brain reserve (``hardware''), cognitive reserve is analogous to the brain's "software" (Stern, 2002), and describes (i) the robustness of a particular cognitive function against brain pathology or (ii) the ability to use alternative functions when a default function is rendered inoperable. Cognitive reserve can be thought to involve the implementation of cognitive processes and representations. Individuals with increased cognitive reserve tend to be more highly educated, possess higher IQs, reach higher occupational attainment, and be involved in a diverse range of leisurely activities. (Stern, 2006). Cognitive reserve is thought to be somewhat independent from brain reserve, based on empirical evidence that individuals with equivalent brain reserve may express variable clinical sequelae as a function of cognitive reserve. The term "cognitive reserve" is thus meant to represent physiological robustness within functional brain networks, while the term "brain reserve" refers to differences in available structural neural substrates (Stern, 2002). See Figure 2 for a traditional threshold model of reserve.

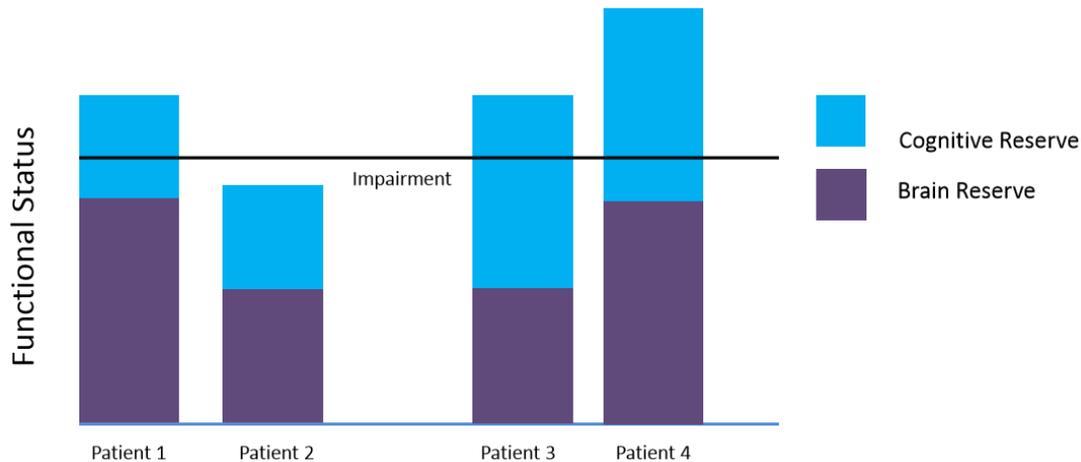

*Figure 2:* A traditional threshold model of reserve (cf. Stern, 2002). Brain and cognitive reserve are represented by measured quantities that cumulatively protect against disease. Patients with greater reserve remain above the impairment threshold following the onset of neuropathology. Patient 1 shows greater resilience to brain pathology than Patient 2 due to greater brain reserve with equivalent cognitive reserve. Patient 3 shows greater resilience to brain pathology than Patient 2 due to greater cognitive reserve with equivalent brain reserve. Patient 4 displays heightened neuroprotection due to the cumulative effects of (i) brain reserve equivalent in magnitude to that observed in Patient 1 and (ii) cognitive reserve equivalent in magnitude to that observed in Patient 3.

Notably, the separation of reserve into a structural and functional component suggests a dualism that does not represent the codependence of these properties in the brain. In a monistic perspective, cognition depends on the function of neurons within the context of the brain's complex spatiotemporal organization (Medaglia et al., 2015). As clinical neuroscience aims to develop innovative treatments for cognitive impairments, an understanding of this organization may inform advanced prediction and treatment strategies (Stam, 2014). Within dynamic network theory, spatiotemporal analysis is fundamental to understanding and intervening in complex systems, and we suggest that this may begin to clarify longstanding problems confronting researchers and clinicians.

**Problems in Traditional Reserve Research**
Three fundamental dilemmas in reserve research challenge future progress in understanding individual variation in cognitive impairments following brain pathology. The first dilemma concerns the potential confusion of statistical prediction and mechanism in clinical cognitive neuroscience. It is common for researchers to posit that brain or cognitive reserve *explains* variance in clinical responses, and therefore forms a fundamental *mechanism* for observed clinical outcomes. However, to date, applied statistical and empirical approaches to reserve may only very weakly be described as addressing *mechanisms;* statistical prediction is necessarily distinct from mechanism discovery (cf. Illari and Williamson, 2012, for an extensive discussion on properties of mechanisms). A mechanism for a phenomenon consists of entities and activities that are responsible for a phenomenon; indeed, the ability to apply knowledge and material sufficient to reproduce a phenomenon may be evidence that a mechanism has been discovered. This standard of knowledge is not frequently available in reserve research. The discovery of fundamental mechanisms of individual differences in resilience to brain pathology is necessary for a mature science of clinical resilience and repair.

The other two dilemmas concern the operationalization of brain reserve and cognitive reserve. Typical approaches to quantifying *brain reserve* involve metrics of the volume of the entire brain (Bigler, 2006; Coffey et al., 2000; Stern, 2006; Sumowski et al., 2013), the volume of specific regions of the brain (Perneczky et al., 2007), lesion loads (Cader et al., 2006), and cerebral metabolism (Cohen et al., 2009). However, empirical studies tend to show weak relationships between these traditionally measured variables and observed clinical status, potentially due to the coarse nature of these variables as gross morphological measures of the brain. In contrast, evidence shows that the patterns of structural connectivity between brain regions are important predictors of brain functional properties (Alstott et al., 2009; Honey et al., 2009; Honey et al., 2010, Becker et al., 2016) and cognitive ability (Medaglia et al., 2015; Wen et al., 2011). These new lines of evidence suggest that employment of advanced network analysis techniques could provide more sensitive measures of the relationship between certain types of brain damage and individual variation in cognitive function.

Finally, approaches to *cognitive reserve* have typically used measures related to life experience and "innate" intelligence (Fratiglioni et al., 2007). Cognitive reserve has been defined as education, occupational achievement (Murray et al., 2011), intelligence quotient (Koenen et al., 2009), leisurely engagement (Scarmeas et al., 2004), and bilingualism (Schweizer et al., 2012). Measures of cognitive reserve have been successful in predicting clinical status in Alzheimer's disease (Stern, 2006), HIV (Foley et al., 2012; Shapiro et al., 2014; Stern, 1996), multiple sclerosis (Booth et al., 2013; Schwartz et al., 2013), normal aging (Sole-Padulles et al., 2009; Tucker & Stern, 2011), stroke (Nunnari et al., 2014; Willis & Hakim, 2013), and traumatic brain injury (Levi et al., 2013). However, the neurophysiological mechanisms that facilitate effective assimilation of experience and development of neuroprotective intellectual abilities over the lifespan remain unclear.

These three dilemmas collectively call for concerted and rejuvenated efforts to uncover neurophysiological mechanisms that link brain morphology and function to cognitive and clinical outcomes. One proposed mechanism of reserve lies in brain plasticity (cf., Fratiglioni, 2007; Mahncke et al., 2006). Processes of structural and synaptic change, functional network plasticity, neurogenesis and vascular development may contribute to brain and cognitive resilience. Stern proposed that it is "the ability to optimize or maximize performance through differential recruitment of brain networks, which perhaps reflects the use of alternate cognitive strategies" (2002). Yet, linking structure, dynamic properties of the brain, and cognitive resilience remains an active area of investigation in need of a framework for integrating structural, functional, and cognitive properties. We propose that substantive progress in research on functional resilience in neuropathological syndromes could result from increasing focus on the nature of dynamic network properties at the macro-level of the human connectome. Importantly, a framework that considers all of these properties simultaneously may be more fruitful than each considered separately. We offer that network control theory provides such a framework.

**Control Theory**

To build the case for dynamic network analysis and the potential utility of network control theory, we first briefly reference foundational concepts from control theory. Control theory is a field of engineering concerned with the analysis of dynamic systems and the design of control algorithms to ensure desirable system behaviors (Karl & Murray, 2008; Kailath, 1980; Khalil, 2002; Levine, 1996; Sontag, 1998). The foundation of control theory can be traced to Maxwell's work examining the dynamics of the centrifugal governor (Maxwell, 1868). A control theoretic representation consists of a dynamic model of the system to be controlled, a reference state or objective that the system should achieve, a controlling mechanism to propel the system toward the reference, and often a feedback mechanism based on the current state that is used to adjust the control signal and manipulate the system in the future (See Figure 3).

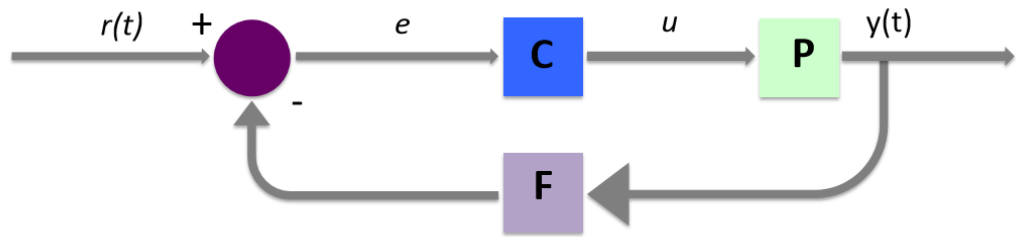

*Figure 3:* A generic classical feedback control scheme. The system is designed to track the reference value r. The output of the system *y(t)* is fed back through a sensor measurement *F* to compare to the reference value *r(t)*. The controller *C* then takes the error *e* (difference) between the reference and the output to change the inputs *u* to the system under control *P*.

For our purposes, control theory can also serve as a template for describing goals, representations, movements, sensory feedback, and the links among these components in neural systems.

Three interrelated core concepts in control theory are critical for our current view: *system identification, observability,* and *controllability.* These concepts concern the ability to describe and manipulate systems. *System identification* refers to the process of building mathematical models of dynamic systems based on observed input-output data. This concept refers to the intersection of real world systems and abstract mathematical models. In principle, there may be a potentially infinite number of models that describe the input-output relationships observed in data equally well (a similar notion to the physical problem of "multiple realizability" in cognitive science (Bickle, 2006)). However, a parsimonious model with equal predictive power to a more complicated model is often desirable in scientific inquiry as well as in practical system design. A *plausible* system identification may be adequate to describe the input-output relationships of a system under a number of conditions, and may be phenomenological in nature. A *valid* system identification is one that accurately and reliably describes the internal state dynamics of a system that mediate between input and output.

The notions of *observability* and *controllability* supplement the practice of parsimonious system identification as a necessary but not sufficient step in full system control. In particular, observability refers to the possibility of reconstructing the system state over time from few and sparse measurements, and without knowledge about the system's initial state. In contrast, controllability refers to the ability to change the internal states of the system via the influence of external input. Observability and controllability can refer to either states or entire systems. The crucial point for our purposes is that observability and controllability are intimately related as mathematical duals (Kalman, 1960), and both relate to the true nature of underlying state transitions.

A control theoretical view of neural systems broadly considers that the nature of cognition and cognitive resilience depend on real neurophysiological processes and states, and that system identification can occur to define a mapping between these properties when due consideration to system observability and controllability is applied. A control view can by flexible applied by the investigator. For example, nervous system is replete with physiological sensing mechanisms that inform the brain about its environment. Modeling in sensorimotor systems has a rich history of mapping the feedforward and feedback mechanisms that relay sensory information and enact control to achieve motor behaviors (Mosconi et al., 2015; Sielder et al., 2004). We suggest that by way of analogy, this strategy can in principle be scaled to consider the entire connectome, where elaboration of the roles of neurons can be identified in control schemes. To build intuition for this case, we next introduce the emerging intersection between control theory and neuroscience.

**Control Theory in Neuroscience**

The application of control theory to neuroscience has already provided critical insights and innovations under an emerging discipline known as *neural control engineering* (Schiff, 2012). This integration between neuroscience and control theory has been developing since the early 2000's (see Voss et al., 2004, for an initial intersection between these disciplines), and has afforded applications for brain-computer interfaces and decoding strategies (Lagang & Srinivasan, 2013; Srinivasan & Brown, 2007) that support adaptive and robust neuroprosthetics (Berger et al., 2011; Gorzelic et al., 2013; Herreros et al., 2014; Hsiao et al., 2013; Taylor et al., 2002). These applications have begun to provide powerful translational opportunities in subcortical systems based on local micro-architectural control models. The interaction between these fields has produced insights into brain systems that serve as feedforward and feedback controllers as well as how to adapt, control, and repair these systems exogenously (see Figure 4 for a basic schematic of a control scheme applied to a system with neural components).

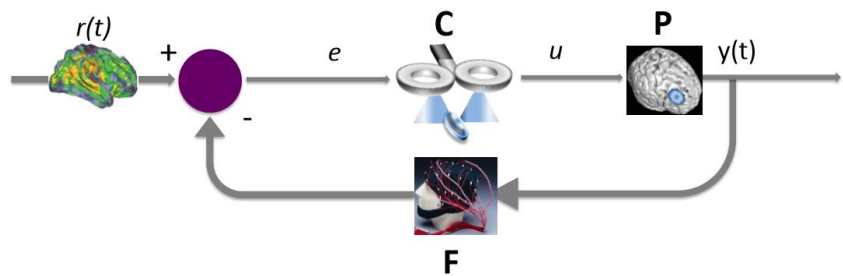

*Figure 4:* A schematic control scheme using a noninvasive brain stimulation technique, transcranial magnetic stimulation. TMS involves the application of a magnetic field in vivo to manipulate neural function via induced current in the cortex (Hallett, 2000). In principle this is one of a broad class of possible stimulation approaches (e.g., microstimulation, deep brain stimulation) in which a control theoretic view can be applied for translational goals. Here the reference r(t) is a particular brain state of interest, which could be an empirically determined state known to be adaptive for a target cognitive process. The controller (C) is a TMS coil that exerts a magnetic field *u* on a target site in the plant, which is the human brain (P). The state output y(t) is read by a feedback sensor (F) such as that sampled by continuous electroencephalographic monitoring. The sensed state is compared to the reference state for the next control iteration. Note that the controller and feedback sensor could be represented by other technologies.

A difficult problem in the domain of control theory is how to quantify and capitalize upon dynamic properties in complex networked systems. We propose that mathematical tools now exist to begin to investigate network control properties of the brain at the level of the connectome in health and in clinical syndromes.

**Neural Dynamics and Network Control**

To facilitate the development of control strategies for neural networks, it is common practice to resort to simplification of dynamic processes in neural circuits, such as linearized generalizations (Galan, 2008) of nonlinear models of cortical circuit activity (see, e.g., Honey et al., 2009). Despite this approximation, the linearized model informs certain controllability properties of the nonlinear representation (Khalil, 2002) and, in fact, control algorithms based on linearized dynamics have successfully been employed for the control of nonlinear systems; see for instance the technique of Gain Scheduling (Leith & Leithead, 2000). Here, we discuss an approach based on simplified linear dynamics in the context of large scale human brain networks to build intuition. Whereas brains are vastly complicated networks with nonlinear dynamics, important intuitions can be gained under some simplifying assumptions. We discuss current and future challenges for full nonlinear control in neural systems at the end of this review.

In the general case, we follow Gu et al. 2015 by defining a generic stationary discrete time network model:

$$(1) \quad x(t+1) = \boldsymbol{A}x(t) + \boldsymbol{B}_K u_K(t)$$

where $x : \mathbb{R}_{\geq 0} \to \mathbb{R}^N$ is the state of brain regions over time and $A \in \mathbb{R}^{N \times N}$ is a symmetric, weighted adjacency matrix with elements $A_{ij}$ indicating for example the number of white matter streamlines (estimated by diffusion tractography) that connect regions *i* and *j*. Elements of $A_{ij}$ are scaled to ensure stable dynamics over a long time interval (Horn & Johnson, 2013). Matrix **B**<sub>K</sub> identifies control points $K = \{k_1,…,k_m\}$ where

$$\boldsymbol{B}_K = [e_{k_1} \ … \ e_{k_m}],$$

such that the vector $e_1$ denotes the *i*-th canonical vector of dimension *n*. Moreover, the input $u_K : \mathbb{R}_{\geq 0} \to \mathbb{R}^m$ indicates the control energy utilized by or applied to the *K* brain regions, and is designed according to the control objective and constraints. We emphasize two perspectives to link network control theory to cognitive resilience and potential avenues for repair. First, we can consider the brain as a system in which different brain regions serve roles in controlling dynamics across the brain. By considering the structure and dynamics required to move the brain among states as a system, properties underlying reserve can be elucidated. Second, the analysis of control inputs in a controllable system can provide information regarding useful strategies for intervention, for example in a brain affected by progressive or acute damage.

**Network Control Theory and Neural Systems**

Extensions of the classical control framework provide the opportunity to study how structural properties of dynamic networks affect their control properties (Rahmani et al., 2009; Rajapakse et al., 2011; Ruths and Ruths, 2014; Pasqualetti et al., 2014). A key mathematical tool to characterize the controllability of a network is the controllability Gramian $\boldsymbol{W}_K$:

$$(2) \ \boldsymbol{W}_K = \sum_{\tau=0}^{\infty} \boldsymbol{A}^\tau \boldsymbol{B}_K \boldsymbol{B}_K^T (\boldsymbol{A}^T)^\tau,$$

where *T* indicates a matrix transpose and $\tau$ indicates the time step of the trajectory. Eigenvalues of $\boldsymbol{W}_K$ measure the degree of controllability in the network. Depending on the assumptions of the model, the structure of $\boldsymbol{W}_K$ can be used to provide guidelines for the optimal control of cognitive functions (Gu et al. 2015).

A network can be controlled in different ways by different types of nodes. That is, based on a node's position in a neural system, it may serve a different control role over the system's dynamics that is defined by the system's structure and dynamics (Gu et al. 2015). We quantify the ability of a network node (in this case, a brain region) to control the rest of the system using one of three control metrics: average controllability, modal controllability, and boundary controllability (Pasqualetti et al., 2014). Regions with high average controllability are efficient in pushing the brain into local easily-reachable states with little effort. Regions with high modal controllability can push the brain into difficult-to-reach states with little effort. Regions with high boundary controllability can push the brain into states in which cognitive systems are either coupled or decoupled. See Supplemental Information and (Pasqualetti et al., 2014) for technical definitions of these metrics.

Henceforth, we simultaneously emphasize that there are two perspectives to apply to nodes in a neural network: either a target to be controlled or the controller. In the first case, we consider that neural systems are increasingly the target of experimental stimulation at the sub-neuron level or at the level of neuronal ensembles. Here, the researcher is interested in understanding how to administer exogenous influences to push the brain into desirable states. Alternatively, we can consider that at a given scale of organization, a node (such as a single neuron or ensemble of neurons) can be modeled as a controller for other components of the system. In this case, the control problem for a neuron or neural ensemble does not represent the goals of the researcher, but the role of the neurons in the context of the nervous system. This does not imply that the neurons have specific goals to attend to; rather, it acknowledges that neurons exert physical influences on one another that collectively govern system behavior. Importantly, a control perspective provides a context to reason about both perspectives simultaneously: understanding how a system naturally controls itself can inform our understanding of how to control the system exogenously.

**Network Controllability and Cognitive Systems**

To draw a link to cognitive reserve research, we now consider recent applications of network control theory in health and disease. The bulk of the new field of neural control engineering has focused on micro-scale and predominantly subcortical systems (Schiff, 2012). At this scale, control theoretic techniques are rapidly influencing our understanding of the dynamics associated with Parkinson's disease and how to use adaptive control via deep brain stimulation to correct motor output. However, how these intuitions might relate to cognition more broadly is not well understood, perhaps due to the challenge of identifying the ideal spatial scale in which to ground the investigation. Whereas some cognitive functions – such as processing visual orientation and spatial frequencies – can be localized to small-scale neural networks (Mazer et a., 2002), other functions such as cognitive control may depend on distributed networks of brain areas (Braun et al., 2015). This larger level of organization appears particularly relevant for understanding brain and cognitive reserve, which depend on complex psychological constructs that involve distributed circuits.

In the first application of network control theory to large-scale neuroimaging data, Gu and colleagues (2015) applied a linear network control model to human structural brain networks. In this work, a critical link between regional controllability and brain system organization was established. The analysis built on high-quality diffusion weighted imaging data acquired in triplicate from eight individuals, and replicated results in macaque cortex and in lower-resolution diffusion tensor imaging acquired in 104 healthy human subjects. The results indicated that average control hubs tend to lie in the default mode system, suggesting that default mode regions are well suited to guide the system into multiple easy to reach states, consistent with the notion that the system represents readiness states prior to engaging in particular tasks (Raichle et al., 2001). Modal control hubs tended to lie in the fronto-parietal and cingulo-opercular cognitive control systems, suggesting that these regions are well suited to guide the system into difficult-to-reach states. This putative control role is consistent with a large body of cognitive neuroscience literature demonstrating their critical role in supporting difficult and complex tasks, (Brass et al., 2005; Dosenbach et al., 2008; Fassbender et al., 2006).

Interestingly, Gu and colleagues also observed that boundary control hubs tended to lie in the ventral and dorsal attention systems. High boundary controllability implies the ability to bridge or separate functional modules, a capability that is critical for learning (Bassett et al., 2011; Bassett et al., 2013), memory (Braun et al. 2015), and cognitive flexibility (Braun et al. 2015). Hubs identified as strong boundary controllers are predicted to be involved in mediating node flexibility across the brain over the course of tasks and during transitions between tasks. This suggests that boundary controllers may draw significant neurophysiological resources when flexible communication is required. Dorsal and ventral attention systems thus are predicted to promote the integration and segregation of distributed cognitive systems, guiding and parsing network resources as necessary to support cognitive tasks (cf. Cate et al., 2012; Thoma & Henson, 2011).

This initial analysis in a healthy sample establishes several foundational principles for the intersection of network control theory and reserve. First, it further supports findings that topological properties of functional brain networks are deeply dependent on gross underlying structural connectivity within the default mode, fronto-parietal, cingulo-opercular, and attention systems (Honey et al., 2010). Second, it establishes that the theoretically predicted preferences of brain regions to perform specific control strategies readily conforms to decades of knowledge in cognitive neuroscience in intuitive and interpretable ways. Third, it establishes a mechanistic framework for connecting brain dynamics to underlying structural network configurations. Finally, and as will henceforth be our focus, network control theory offers analytical and computational approaches that advance network neuroscience and inform its use in the study of dysfunction in neuropathological syndromes.

**Brain Structure, Dynamics, and Neuroplasticity**
*Examining the Brain as a Dynamic System via Network Control Theory*

Several theoretical implications should be clarified to establish network controllability as a critical concept in research involving cognitive reserve in neuropathological populations. As opposed to the distinction between brain reserve (structure) and cognitive reserve (function) in clinical cognitive neuroscience, network control theory enables us to explicitly link structure and function by describing properties of the human structural connectome that develop over time and support the dynamics of cognition. Indeed, this conceptual paradigm enables us to link the otherwise divergent notions of brain structure, network topology, control capabilities, neuroplasticity, and cognitive function at a point in time when the diverse fields of study they draw from are becoming sufficiently mature to investigate initial hypotheses.

An overarching schematic of the intersection of network controllability and brain function can be seen in Figure 5.

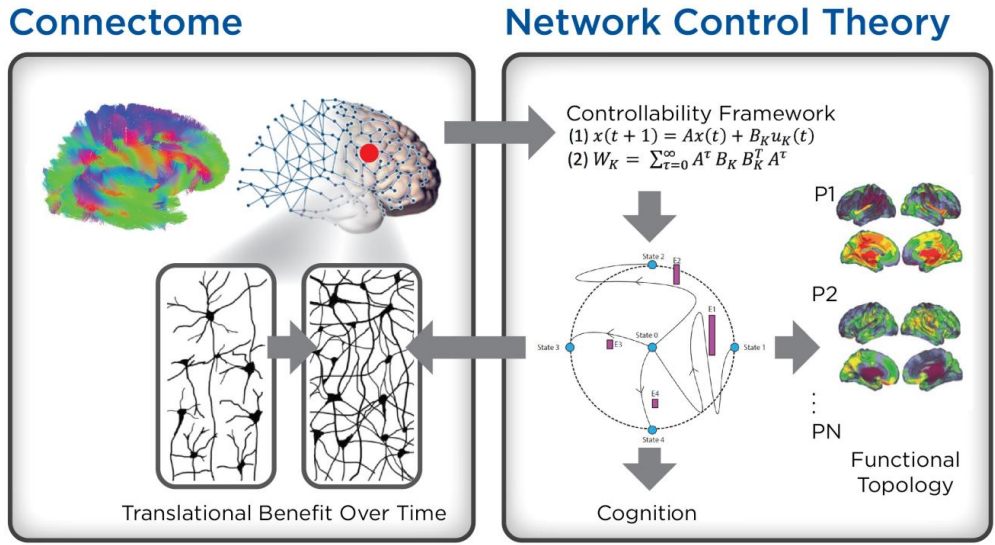

*Figure 5:* Network Control Theory and the Connectome. Network control theory is a means for expressing the dynamic properties of the brain that mediate between structural networks and the functional properties they afford. Traditional reserve measures such as global brain size, regional size, and developmental variables (left panel) have influences on cognition via their underlying structural network configurations. In turn, structural network configurations plus local signaling dynamics establish the controllability of the brain system (center panel). Variance in network controllability implies variance in functional topologies such as those computed from functional magnetic resonance imaging (fMRI), electroencephalography, or electrocorticography. It also implies qualitatively different dynamics that support cognition. The controllability of regions results in neuroplastic changes due to controlling the flow of neural signaling cascades in addition to momentary maintenance and switching of cognition and cognitive modes. Neuroplasticity in turn exhibits feedforward influences on controllability over time.

This intersection begins by addressing brain structure from a traditional reserve perspective in which brain volume is used as a proxy of total neural mass and regional volume for local neural mass. From a network control perspective, brain volume and regional volumes are important to the extent that they represent neural tissue configured to support optimal control characteristics of brain regions. That is a theory of network control in clinical neuroscience is explicitly concerned with structure insofar as it relates to the brain's functional dynamics and cognition, directly encouraging a shift from descriptive correlations to predictive mechanisms. Such a theory builds on the now well-developed lines of study predicting resting state functional networks (Seeley et al. 2007) from structural networks (Breakspear et al., 2007; Cassiano et al., 2015), predicting cognitive function from network architecture (Medaglia et al., 2015), and predicting the effects of damage to structural networks on functional resting networks (Alstott et al., 2009). But this theory also offers a new perspective, by predicting which regions of the brain are suitable drivers of cognitive dynamics, acknowledging the system-wide functional affordances of a brain region depend in part on its connectivity with the rest of the system. Critically, these predictions cannot be derived from brain volumetrics, functional topology investigations, static structural network analysis, and behavioral techniques in isolation.

We suggest that for initial applications in cognitive reserve, in line with Gu and colleagues (2015), the connectome can be sampled at the level of resolution afforded by modern diffusion tractography to examine the macro scale organization of the system and the role of individual regions in the system. By formulating the problem of cognitive resilience and failure as a network control problem, the researcher can estimate a brain region's predicted potential to drive the brain toward multiple easily reachable states (average controller), difficult to reach states (modal controller), and states of system segregation or integration (boundary controller). States are reached by the time-dependent influence of one region on other regions to which it is connected, emphasizing that controllability is a function of a region's local network neighborhood. These dynamic interactions are thought to support general cognitive processes as well as processes associated with cognitive reserve (e.g., IQ, bilingualism, education). Importantly, the control roles of nodes in the network can be validated directly against measured functional neuroimaging and other physiological measurements. As the nature of the brain's structure and dynamics are clarified, links between the controllable brain and measures from techniques such as fMRI, electrocorticography, electroencephalography, and magnetoencephalography in conditions of rest and cognitive tasks may be discovered. In this view, the dualism of brain and cognitive reserve are considered in one framework, where the term "reserve" refers to the joint structural and functional characteristics of brain networks that offer cognitive protection in disease. Concretely, the combination of structural connections among regions in brain networks ("edges" in a graph representation) and the dynamics mediated by them form the physical basis of reserve.

Finally, network control theory offers a novel way to frame notions of neuroplasticity. In particular, rather than focusing on individual synapses or on the functional recruitment of particular regions, it may behoove us to discuss the dynamic influence of a node on the system For example, distributed neural activity may entrain synaptic processes, resulting in widespread neuroplastic changes over time, that in turn modify the controllability profile of the brain in a feedforward manner (See Figure 5; see also "Nonlinearity, Multiple Scales, and the Time-Varying Brain). The consequences of damage to controllers of different classes might then lead to variable dynamic entrainment of synaptic processes over time. In principle, such a process could explain why different patients exhibit different functional profiles in neuroimaging data that underlie varying learning and recovery trajectories in neuropathological syndromes.

In sum, in the network control view, if brain morphological features such as total or regional size are predictors of clinical status, it is because they are proxies for underlying structural configurations with functional controllability affordances. For example, if a given brain has a rich repertoire of intact modal controllers, it may be more likely to easily adapt to a range of difficult challenges to the network. That is, conditions involving novelty, increased cognitive load, or general network damage may be more easily supported by a connectome that can more easily drive the system into difficult to reach states to manage challenges. This would represent dynamic network mechanisms underlying reliable observations that challenge-responsive regions remain responsive to challenge in disease states (Hillary, 2008). Intact average controllers may support routine operations and flexibility to engage in a range of tasks from other states of engagement or relative disengagement from the environment. These regions and their dynamic interactions may support the brain's efficient recruitment of learned procedural and declarative information to manage a wide array of needs. Intact boundary controllers may support the flexible integration or segregation of activity across brain networks. This may mediate the brain's ability to maintain focus, flexibly and quickly adapt to new circumstances, integrate across multiple sensory modalities, and learn efficiently.

*The Translational Appeal of Network Control Theory*

Adoption of the network control perspective could increase opportunities to prognosticate individual differences in clinical outcomes and identify opportunities for intervention that are not available under current theoretical paradigms. By leveraging rapidly developing advances in network science, network control theory provides a formal integration of the study of the connectome and dynamic systems approaches. As variations in network controllability are investigated in normalcy and disease, the relationship between controllability distributions and individual differences in resilience and recovery from injury can be investigated. Increasing focus on classification of network dynamics underlying disease presentations could advance theoretical models of disease in line with current agency-funded initiatives (e.g., the National Institutes of Mental Health Research Domain Criteria initiative; Insel et al., 2010). Specifically, such investigations could establish models of disease defined by the dynamic consequences of diseases in networks: clinical phenotypes can be considered to be the result of failures in brain dynamics that depend on structural and dynamic profiles. Examining the dynamic profile and trajectories of the evolution of cognitive changes in neuropathology can become an active program of research with a central goal toward clarifying principles of dynamic function, which is directly related to the ability to predict, control, and repair the system.

**Specific Hypotheses of Network Control Theory in the Brain**

Several hypotheses based on network control theory could provide a basis for empirical investigation in reserve research. The challenge is to provide plausible hypotheses from a network control perspective for network function and resilience. As described above, the suggestion is that dynamic network properties are fundamentally determined by network control properties, which distinguishes the current hypotheses from previous treatments. Here we describe several hypotheses designed with the explicit goal of providing a unification of various approaches to brain and cognitive reserve under the framework of network control.

- *Hypothesis 1: Brain and regional volume are partial proxies for network and node controllability*

To encourage productive connections to traditional approaches in reserve research, we suggest that previous findings in neuroimaging measures are important to the extent that they provide predictive power. We further suggest that these measurements are likely related to network control properties including robustness to failure or resilience to perturbations. However, in truth, little is known about the relationship between (i) the structural configuration of networks in the connectome and (ii) raw and regional brain size, or how this relationship might vary across development, or in healthy *versus* disease states. Controllability analyses will provide a basis for extension of traditional volumetrics into network neuroscience. Whether controllability characteristics of the brain are predicted by volume would establish the first link between network control theory and reserve, and is arguably expected given the deep dependence of controllability on underlying network structure.

Importantly, it is likely the case that the reserve afforded by volumetric properties is not completely statistically mediated by structural configuration characteristics at the meso- and macro- scale. In principle, all functional properties that afford functional resilience are physically mediated through network architecture. However, some components of brain resilience at the level of networked cytoarchitecture are inaccessible to the resolution of modern neuroimaging techniques. In light of this, a competing hypothesis within brain reserve could be proposed. Specifically, it could be anticipated that if gross or regional brain volume represents the quantity of neurons and synaptic densities, and these features themselves genuinely represent a form of brain reserve, then the networked structural properties accessible to neuroimaging tractography techniques will be incomplete and partial predictors of clinical status. In this case, traditionally used imaging measures should explain unique variance that is not accounted for by meso- and macro-scale network architecture.

- *Hypothesis 2: Education, genetics, nurturing, and nutrition will predict controllability profiles via influences on network structure*

Environmental influences are critical to brain structural development (Blakemore & Cooper, 1970; Giedd et al., 2014; Petanjek et al., 2012). As described under Hypothesis 1, the influence of environmental effects on structural network development is comparatively unknown. However, we anticipate that early lifespan environmental influences will at least partially determine important variance in underlying structural networks, and therefore bear consequences for global and regional network controllability. Further, we anticipate that network controllability characteristics will have feedback influences on gross structural development in early life and lower-scale structural changes via known neuroplastic processes throughout life (Gong & Van Leeuwen, 2004; Katz & Shatz, 1996; Kwok et al., 2007; Rubinov et al., 2009). Thus, the distinction between the reserve view and a network control view is that environmental and developmental effects are important to the extent that they produce intact and robust network controllers.

- *Hypothesis 3: Research in brain network controllability will directly inform neural system models and treatments*

Just as a dynamic model is fundamental to network control theory, so it may be critical for intervention. The different control roles identified in previous work imply different responses to noninvasive brain stimulation (Muldoon et al., 2016). If opportunities to prospectively influence the brain toward desirable states are available as a result of dynamic models, the link to informative translational approaches will be a matter of model refinement at the intersection of biophysics and cognitive science. Once the states and trajectories corresponding to optimal individualized cognitive performance, optimal control strategies (cf., Betzel et al., 2016) to intervene in pathological states can be considered in real world contexts.

**Nonlinearity, Multiple Scales, and the Time-Varying Brain**

The study of network dynamics underlying cognitive function is in its infancy in the cognitive neurosciences, perhaps especially in modern neuroimaging. Thus far, we have focused on recent work that makes simplifying assumptions about brain structural and dynamic analysis for heuristic appeal. Before cognition is ever examined, nontrivial challenges confront the brain-behavior scientist if a truly integrated translational science is the goal. These are especially salient in the context of network control theory. To realize a science of network control in the context of cognition and reserve research at multiple levels of brain organization, observations from several lines of research should be considered and integrated. We briefly highlight several relevant themes in this area. Specifically, we note that brain systems exhibit non-linear (Schiff et al., 2012), non-stationary (Bassett et al., 2011; Bassett et al., 2013; Gonzalez Andino et al., 2001; Jones et al., 2012; Snider & Bonds, 1998; Tomko & Crapper, 1974; Zhan et al., 2006), and thus non-ergodic (Bianco et al., 2007; Medaglia et al., 2011; Werner, 2011) dynamic properties on multiple scales of organization.

Brains exhibit nonlinear dynamics at all scales of organization (Schiff et al., 2012). As discussed, network control theory has already revealed intuitive associations between dynamic statistics in a linear example of network control theory and the organization of cognitive systems (see Gu et al., 2015, supplemental discussion). This work has recent extensions into the prediction of the nonlinear effects of noninvasive brain stimulation (Muldoon et al., 2016; Spiegler et al., 2016) and changes in optimal control trajectories following node loss (Betzel et al., 2016). However, it will be crucial to develop refined solutions in neural system identification to encourage optimal theoretical models and practical control-based interventions. At the scale of organization considered here, hemodynamic models (Friston et al., 2000) and dynamic causal models that rely on them (Friston et al., 2003) based on ordinary differential equations have proven successful in low-dimensional systems. More recently, local nonlinear oscillation models of region dynamics (Wilson & Cowan, 1973) in macro-scale structural networks have yielded successful predictions of functional correlations and coherence among regions in neuroimaging data (Cabral et al., 2011). Depending on the scale and control goal, existing nonlinear approaches may prove useful, and should be evaluated against measured functional signals and system behaviors (i.e., cognitive function and behavior) induced by control actions such as brain stimulation. While nonlinearity and high dimensionality are often considered obstacles to control, recent mathematical and computational approaches are rapidly addressing these limitations (Motter, 2015).

Brains exhibit non-stationary dynamics. Brain dynamics change stochastically (Kelso et al., 1997) and across cognitive conditions (Medaglia et al., 2015) even when most observable features of the brain's structure remain relatively stationary (though this is more a matter of spatiotemporal scale than presence or absence of non-stationarity). Non-stationarity in brain activity demonstrates that it is weakly non-ergodic: the average process parameters over time and the entire neural ensemble are not the same. All possible states are accessible, but some require very long times to visit (Bianco et al., 2007). This is a consequence of the complexity of meso- and micro-scale neural dynamics underlying measurable neuroimaging signals. Even more challengingly, at least part of brain non-stationarity on middle and long-term time scales depends on neuroplasticity in the context of health and neuropathology. For a complete science of network control theory, explicit modeling of the sources of non-stationary dynamics and how they interact with neuroplasticity may reveal relationships between these two fundamental aspects of brain function and will be necessary for the design of optimal control strategies.

Finally, brains have a complex modular multiscale organization (Bassett et al., 2011). Research in this area is in its promising infancy in cognitive network neuroscience (Medaglia et al., 2015), and presumably lies at the heart of the brain's complex dynamic and cognitive characteristics. While early evidence suggests promise in the analysis of simplifying assumptions about network controllability at the macro scale of organization in the human brain, substantial progress will be necessary across many levels (Kopell, 2014) to understand and resolve the clinical challenges that confront us. We suggest that in tandem with important developments in subcortical systems (Schiff, 2012), it is equally important to consider much higher scales of neural network organization in the study of cognitive resilience and repair. As it has been suggested that all networked dynamical systems can be considered in a control theoretic view (DeVille & Lerman, 2013), it is possible that fundamental properties that inform system analysis and manipulation can be discovered in the human brain.

**Practical Challenges in Cognitive Neuroscience and Reserve Research**

If anything is clear in modern clinical neuroscience research, it is that successful translational innovation will be field-bridging, multidisciplinary, and collaborative. To achieve the translational goals of prediction, control, and repair in a network control theoretic view of the clinical neurosciences, expertise in biophysics, statistical physics, computation, engineering, and psychology will be essential. To achieve such goals, motivated investigators must work in teams of increasing diversity and complexity. This will initially present technical, linguistic, and cultural barriers to overcome. A shared goal to understand, control, and repair neural systems will be essential to success. Additionally, as noted by at least one author in the study of smaller scale neural systems, our models of neural dynamics are wrong, our measurements are bad, and our computers are never fast enough (Schiff, 2012). This bears substantial implications for the ability to observe and control real neural systems. However, persistent collaborative work on each of these forefronts suggest that the situation will only improve. Our success in developing targeted control theory informed interventions will improve with sustained and novel collaborations between theoretical, technical, and applied disciplines. In sum, the clinical neurosciences may benefit from focused work at a novel scientific intersection between three core themes discussed here (Figure 6).

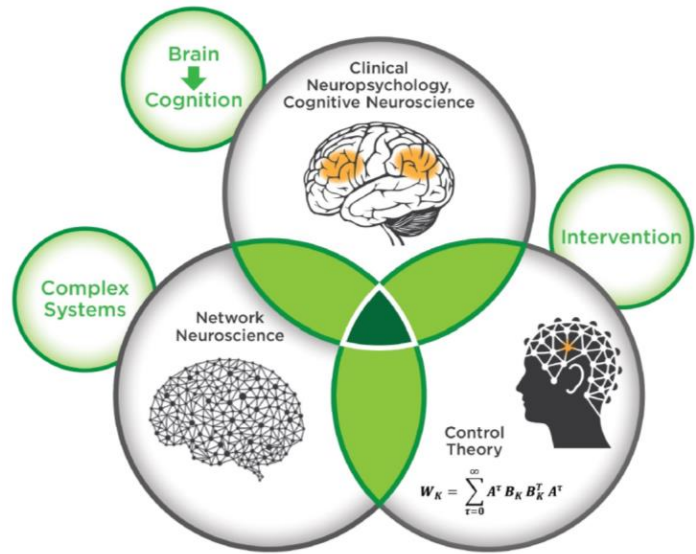

*Figure 6:* A novel scientific intersection. Computational, theoretical, and practical approaches can contribute to a network control based approach to brain and cognitive research for translational benefits. The strengths of each discipline complement the limitations of the others. Cognitive neuroscience and neuropsychology provide models for cognition and examine associations between the brain and behavior. Network science allows us to describe and explain nonlocal complexity in neural systems. Control theory gives us the means to identify control roles and strategies in the neural data. The overlapping areas between each pair of fields represents potential subdisciplines in early stages of development, and the overlap among all three identifies a novel domain focused on control theory-based translational neuroscience.

**Conclusion**

Network control theory provides mathematical tools to assess the control role of elements of the brain based upon simple network parcellation schemes with simplified network dynamics. This approach has been successful at providing a preliminary mechanistic basis for the different functional topologies of known cognitive systems (Gu et al., 2015). We have provided an optimistic case that a network control framework provides the opportunity to promote a union between dynamic network systems approaches and dilemmas in cognitive and brain reserve that can extend beyond traditional approaches. Implementation of network control strategies suggested here in neurally compromised populations will test our initial hypotheses and provide an opportunity to link fundamental problems in reserve theory with implications for translational engineering interventions.

More broadly, formally representing the brain as a dynamic network with an explicit focus on integration across the sciences may lead to the discovery of fundamental theories unavailable within current specific paradigms. Enlightenment could result from this framework through three pathways in the cognitive and clinical neurosciences. The first may result in the discovery of fundamental principles that govern function and emergence in dynamic networks in general. The second may result in the discovery of relationships between features of the dynamic network representation and classically observed cognitive, emotional, behavioral, and pathological features in humans. The third may result in the discovery of new features in dynamic systems that were neither predicted in mathematical formulations nor available via traditional approaches to the neurosciences. These possibilities can begin to be explored now in the human neurosciences and may result in crucial exchanges with developments in dynamic network theory for the foreseeable future. We encourage the reader to pursue such opportunities.

**Acknowledgements**
The authors thank Kathryn R. Hall for her generous contributions to Figure 5 and Figure 6.

**Funding Sources**
This work was supported from the John D. and Catherine T. MacArthur Foundation, the Alfred P. Sloan Foundation, the Army Research Laboratory and the Army Research Office through contract numbers W911NF-10-2-0022 and W911NF-14-1- 0679, the National Institute of Mental Health (2-R01- DC-009209-11), the National Institute of Child Health and Human Development (1R01HD086888-01), the Office of Naval Research, and the National Science Foundation (BCS-1441502 and BCS-1430087). FP acknowledges support from BCS-1430280. JDM acknowledges support from the Office of the Director at the National Institutes of Health (1DP5OD021352-01).